# Two-Point Statistics of Coherent Structure in Turbulent Flow

**Jun Chen** 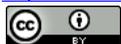

State Key Laboratory for Turbulence and Complex Systems, Department of Mechanics and Engineering Science, College of Engineering, Peking University, Beijing, China
Email: jun@pku.edu.cn

## Abstract

This review summarizes the coherent structures (CS) based on two-point correlations and their applications, with a focus on the interpretation of statistic CS and their characteristics. We review studies on this topic, which have attracted attention in recent years, highlighting improvements, expansions, and promising future directions for two-point statistics of CS in turbulent flow. The CS is one of typical structures of turbulent flow, transporting energy from large-scale to small-scale structures. To investigate the CS in turbulent flow, a large amount of two-point correlation techniques for CS identification and visualization have been, and are currently being, intensively studied by researchers. Two-point correlations with examples and comparisons between different methods are briefly reviewed at first. Some of the uses of correlations in both Eulerian and Lagrangian frames of reference to obtain their properties at consecutive spatial locations and time events are surveyed. Two-point correlations, involving space-time correlations, two-point spatial correlations, and cross correlations, as essential to theories and models of turbulence and for the analyses of experimental and numerical turbulence data are then discussed. The velocity-vorticity correlation structure (VVCS) as one of the statistical CS based on two-point correlations is reiterated in detail. Finally, we summarize the current understanding of two-point correlations of turbulence and conclude with future issues for this field.

## Keywords

Two-Point Correlation, Coherent Structure, Boundary Layer, Wall-Bounded Flow, Turbulent Shear Flow

## 1. Introduction

The common recognition of the existence of organized motions or vortices in turbulent shear flows can be traced to the works of Theodorsen (1952) [1] and Townsend (1956) [2] over six decades ago. In the past 60 years, a great deal of insight into the characteristics of organized structures in turbulent shear flows has been achieved, primarily by means of flow visualization techniques [3] [4] [5] [6] [7]. Some combined flow visualization and quantitative techniques [8] [9] [10] [11] have demonstrated the significance of certain events, e.g. the bursting process, or structures in turbulence.

Unfortunately, the present knowledge of organized motions has seldom been applied to turbulence theories or to quantitative turbulence models. This is partially caused by the lack of a quantitative definition of organized structure and an objective method to assess their contributions to turbulence stresses, especially their role in producing turbulence. Additionally, most flow visualizations have been carried out at low Reynolds numbers, in which the limited complex of turbulence makes it easier to detect organized motions. Thus, much of our knowledge of coherent motions is limited to the structures that can be observed in flow visualizations. It is extremely desirable to have the method for extracting coherent structures (CS) from turbulent flow fields and to evaluate their contributions to turbulence statistics, disregarding how turbulent the flows are.

Quantitative descriptions of organized structures are the essential element for successfully applying the knowledge of structures to engineering models, and the need of them has led to the use of statistical techniques. Most statistical techniques for identification of organized structures from turbulent flows will nearly make a structure from any stochastic field, regardless of their existence in the field. Therefore, the connection of statistically derived structures with important instantaneous events must be asserted based on independent knowledge of the relevant dynamics. Thus, the result of such statistical techniques is an ensemble-averaged structure. This statistical structure is often confined to a relatively small portion of the flow domain, with the surrounding fluctuations being averaged out. For that reason, the inherent symmetries in the statistics naturally impose symmetries in the statistics naturally impose symmetries on the extracted structures. Hence, the structures obtained from statistical techniques represent the average of the most energetic events and this is helpful for understanding the mechanisms of turbulence.

In reality, whether a statistical structure is truly found in the instantaneous flow field, and on what condition it exists, may not be the critical question. If the main objective of a study of coherent motions is to find a decomposition of turbulence into deterministic and stochastic parts, identifying the statistical structure is certainly an efficient technique. In addition, the coefficients of some statistical structures can be directly derived from the Navier-Stokes equations, which are of fundamental significance for theoretical development. Here, we emphasize that for modeling purpose, details of the instantaneous structures are not of interest to us, but the statistical structure may actually be what is important.

An interpretation of a CS is a fluid mass connected with the *phase-correlated vorticity* [12]. In determining the phase of fluids, two- or multi-point measure-ments of turbulence fields are common techniques applied in experiments. The two-point correlation tensor was found capable of accurately estimating conditional velocity fields [13]. Thus, the conditionally averaged velocity can be obtained from the information contained in the velocity correlation tensor. Some of the possibilities for deducing the structure of complicated flows from simultaneous measurements at multiple points have been demonstrated in experiments on turbulent boundary layer (TBL) [14].

Different kinds of two-point correlation might be used, e.g. velocity-velocity, velocity-vorticity, or conditioned correlation, as reported by Sillero and Jiménez (2014) [15], Chen *et al.* (2014) [16] and Hwang *et al.* (2016) [17], according to the original intention of the study. These techniques are often used to characterize the behavior of the coherent motions in turbulent flows, *i.e.* the shape, size and location of CS. The concept of two-point correlation in determining the statistical properties of turbulent flows has been used in constructing turbulence theories. In inhomogeneous equilibrium flows the logarithmic law is the groundwork in statistical turbulence theory. The log-law has been found in a broad range of different turbulent wall shear flows, and several statistical models have been proposed to be consistent with the log-law. Particularly, the two-point statistics of the log-layer has been applied to develop the rapid distortion theory (RDT) [18], which can be applied to solve unsteady flows where the non-linear turbulence-turbulence interaction can be neglected in comparison to linear turbulence-mean interaction.

This article highlights some of the major developments in applying two-point correlations to the study of turbulence. The paper is organized in the following way. Section 2 reviews two-point correlation techniques used to investigate turbulent shear flow. Section 3 evaluates a statistical structure based on two-point cross-correlation, velocity-vorticity correlation structure (VVCS), in wall-bounded turbulence, and assesses the compressibility effects and the flows in special boundary conditions. Finally, Section 4 concludes with future perspectives.

## 2. Two-Point Correlation Coefficients

For many decades, two-point correlations have been a milestone of statistical theories of turbulence in modeling of its processes, and they have become virtually essential methods of data analysis for investigating turbulent flows. As one of these methods, the Eulerian correlation coefficient of velocity components in stationary turbulence, fluctuation about their mean values, is defined for two locations and two times as

$$R_E\left(\boldsymbol{x}, \Delta\boldsymbol{x}, \tau\right) = \frac{\left\langle u_i\left(\boldsymbol{x}, t_0\right) u_j\left(\boldsymbol{x} + \Delta\boldsymbol{x}, t_0 + \tau\right)\right\rangle}{\sqrt{\left\langle u_i^2\left(\boldsymbol{x}, t_0\right)\right\rangle}\sqrt{\left\langle u_j^2\left(\boldsymbol{x} + \Delta\boldsymbol{x}, t_0 + \tau\right)\right\rangle}} \tag{1}$$

where $u_i$ and $u_j$ $\left(i, j = 1, 2, 3\right)$ denote the velocity fluctuations; $\boldsymbol{x} = \left(x_1, x_2, x_3\right)$ is a specified reference location; $\boldsymbol{x} + \Delta\boldsymbol{x} = \left(x_1 + \Delta x_1, x_2 + \Delta x_2, x_3 + \Delta x_3\right)$ are locations with respect to $\boldsymbol{x}$ that can be consistently varied, and $\boldsymbol{\tau}$ is the time in-crement between the two times, $t_0$ and $t_0 + \tau$. Here, the numbered indices, $i$ and $j$ $(i, j = 1, 2, 3)$, respectively indicate the streamwise ($x$-direction), wall normal ($y$-direction) and spanwise ($z$-direction) directions, and $\langle \cdot \rangle$ denotes an ensemble average of realizations.

Another correlation coefficient can be expressed as Lagrangian correlation coefficient, which is defined for properties of fluid particles that pass through Eulerian locations $x$ at time $t_0$ and travel along Lagrangian trajectories to arrive at positions $x + \Delta x(\tau)$ at time $t_0 + \tau$. In this case, the displacement vector, $\Delta x(t_0 + \tau)$, is a random variable representing the positions, at time $t_0 + \tau$, of the particles in the averaging ensemble with respect to the initial location $x$ at time $t_0$. Thus, for Lagrangian correlation coefficients, $\Delta x$ and $\tau$ are interdependent, *i.e.*, $\Delta x$ is a function of $\tau$. Hence, Lagrangian correlation coefficients are given by

$$R_L(x, \tau) = \frac{\langle u_i(x, t_0) u_j(x + \Delta x(t_0 + \tau)) \rangle}{\sqrt{\langle u_i^2(x, t_0) \rangle} \sqrt{\langle u_j^2(x + \Delta x(t_0 + \tau)) \rangle}} \tag{2}$$

Two-point spatial correlation of two variables (also called *two-point cross-correlation*) is a procedure usually applied to investigate the spatial relation between two variables, respectively, at two spatial points in a turbulence field. Two-point cross-correlation coefficient of two variables $\phi_i$ and $\psi_j$ is defined for two locations as

$$R_{ij}(x_r, x) = \frac{\langle \phi_i(x_r) \psi_j(x) \rangle}{\sqrt{\langle \phi_i^2(x_r) \rangle} \sqrt{\langle \psi_j^2(x) \rangle}} \tag{3}$$

where the subscription $r$ denotes the reference point; $\phi$ and $\psi$ are fluctuations of two variables or components, e.g. velocity, vorticity, temperature, pressure, shear stress. Two- and three-dimensional representations of correlation function of two variables with respect to the reference point always give useful information about the structure of the flow variables. Moreover, the correlations of the different variable components may have very different geometries [16].

## 2.1. Space-Time Correlation

Autocorrelation is a characteristic of data which shows the degree of similarity between the signals of single variable over consecutive time intervals. The significance of the autocorrelation lies in the fact that it indicates the memory of the process. The space-time autocorrelation (also called space-time correlation), as a type of two-point correlation, can be defined with the Eulerian (Equation (1)) and Lagrangian (Equation (2)) frames of reference. As a common method for exploring the coupling of spatial and temporal scales of motion in turbulence, it has been employed to develop time-accurate turbulence models for the large-eddy simulation of turbulence-generated noise and particle-laden turbulence. Further details of the knowledge of space-time correlations and the future issues for the field have been reported in [19].

A typical model constructed by using Eulerian two-point two-time correlation is the so-called "elliptic model" in analyzing turbulent shear flows with a second order approximation to the iso-correlation contours [20]. He and Zhang (2006) suggested that the space-time correlations of velocity fluctuations are mainly determined by their space correlations and the convection and sweeping velocities, and developed the elliptic model, relating the space-time correlations to the space correlations through the convection velocity and the sweeping velocity. The model was validated by its agreement of the normalized correlation curves from the direct numerical simulation (DNS) of turbulent channel flows.

Space-time correlations of fluctuating velocities in terms of the elliptical model were lately examined in numerically simulated turbulent shear flows [21]. The elliptic model was extended to the inertial range based on the similarity assumptions of the iso-correlation contours. The analytical expressions for the convection and random sweeping velocities were derived from the Navier-Stokes equations for homogeneous turbulent shear flows, and a universal form of the space-time correlations with the two characteristic velocities was then obtained.

Eulerian two-point two-time correlations, as expressed in Equation (1), are often employed to obtain the convection and propagation processes, such as ejections, sweeps, and shearing in TBL. The propagation speed of turbulent fluctuations was first suggested in Taylor's hypothesis on isotropic turbulence. Taylor postulated that the spatial fluctuations can be inferred from temporal signals of flow variables by assuming frozen turbulence—the fluctuations at $(x, t)$ behave as $(x - U_c t, 0)$, where $U_c$ is the propagation speed of the fluctuations. In the past several decades, propagation speed has received considerable attention in the study of turbulent shear flows because of the presence of large-scale organized motions, e.g. CS. The inherent nonlinear and nonlocal nature of the interaction between flow structures and compressibility bring more difficulties in understanding the propagation process of fluctuations at high speeds. Despite much efforts, a quantitative interpretation of the propagation speed at all distances from the wall has not been achieved.

Earlier speculation of the near-wall propagation being induced by advection of coherent vortex structures in TBL was presented in [22]. Krogstad, Kaspersen, and Rimestad (1988) reported in their experiments showed that the convection velocities are constant in the buffer layer, whereas remarkable variations in $U_c$ occur in other regions, both with regard to the distance from the wall and the scale of the motion [23]. It is also shown that the largest-scale motions are convected at velocities close to the local mean velocity. Whereas, the velocities drop significantly as the scales are reduced for all types of events studied. Cao *et al.* (2008) found two characteristic convection speeds near the wall in turbulent channel flow—one associated with small-scale streaks of a lower speed and another with streamwise vortices and hairpin vortices of a higher speed [24]. This flow structure illustrates the dominant role of CS in the near-wall convection, and the nature of the convection. The propagation of patterns of velocity fluctuations is scale-dependent. Thus, the structure convection velocity should be associated with the evolving state of the structure.

The aforementioned facts were lately proved in DNS of compressible turbulent channel flows. Pei *et al.* (2012) proposed a model for the entire profile of the propagation speed of the streamwise velocity fluctuations, $U_c(y)$, for incompressible and compressible wall-bounded turbulence [25], which was examined in numerically simulated turbulent channel flows at four Mach numbers ($Ma = 0 \sim 3.0$). Eulerian two-point two-time correlation was used in the study to obtain profiles of $U_c$, which showed remarkable similarity at different Mach numbers.

## 2.2. Two-Point Spatial Correlations

The concept of autocorrelations can be extended to multi-point statistics. Consider for example, the correlation between the velocity at one point and that at another. If the time dependence is suppressed, this correlation is a function only of the separation of the two points. In fact, two-point spatial correlation coefficient of velocity has been widely used to measure the size of CS in complex turbulent flows. Note that these correlations are high-dimensional quantities. The correlation for channel flows is originally four-dimensional, while it is five-dimensional in boundary layers with homogeneity in the spanwise direction [26].

Afzal (1983) conducted DNS of the boundary layer subjected to strong adverse pressure gradient at Reynolds number $Re_\theta = 150$ - $2200$ to investigate the features of TBL under adverse pressure gradient (APG) [27]. The two-point spatial correlation $R_{uu}$ obtained far away from the transition region at $Re_\theta = 2175$ and at $y/\delta = 0.4$ indicated that the correlation length of $u$ in an APG boundary layer is virtually half that in the ZPG TBL. The wall-normal and spanwise correlations, $R_{vv}$ and $R_{ww}$, are quite different from $R_{uu}$. The length scales are more limited and the correlated regions are elongated in the wall-normal direction, as observed in ZPG boundary layers.

Two-point spatial correlations were also employed to investigate the axisymmetric turbulent flows. Since the theory of axisymmetric turbulence was first developed by Batchelor (1946) [28], a few scholars developed axisymmetric tensor theory of axisymmetric turbulence. Kerschen (1983) derived the constraints for the two invariant functions which appear in the expression for the two-point velocity correlation tensor [29]. It is noteworthy that only limited progress has been made on the theory of axisymmetric turbulence. For example, relations between longitudinal and lateral correlations and spectra are not available. This would hinder the applicability of the theory. Later, Zhu *et al.* (1996) derived a relation between longitudinal and lateral correlation functions and tested it with available data [30].

Two-point velocity correlation functions and integral length scales have extended its application to complex flows. Alexander and Hamlington (2015) reported detailed statistical measures of the turbulent environment vertical profiles of Reynolds stresses, two-point velocity correlations, and velocity structure functions for understanding of localized loading of an ocean current turbine, in or-der to accurately predict turbine performance and durability [31]. Large eddy simulations (LES) of tidal boundary layers without turbines were used to measure the turbulent bending moments. The impacts of waves and tidal velocity, boundary layer stability and wind speeds on the strength of bending moments were analyzed. The results showed that both transverse velocity structure functions and two-point transverse velocity spatial correlations can be used to predict and understand turbulent bending moments in tidal channels.

Nowadays two-point velocity correlation function is a popular approach to understanding turbulent BL. Ganapathisubramani *et al.* (2005) conducted stereoscopic particle image velocimetry (PIV) measurements in streamwise-spanwise and inclined cross-stream of a TBL at moderate Reynolds number ( $Re_\tau \sim 1100$ ) [32]. Two-point spatial velocity correlations computed with a feature-detection algorithm in the log-law region using the PIV data was used to identify packets of hairpin vortices. Both streamwise-streamwise, $R_{uu}$, (see **Figure 1**) and streamwise-wall-normal, $R_{uw}$, correlations present significant spatial coherence in the streamwise direction, such that the correlation coefficient extends further than 1500 wall units ($\sim 1.5\delta$). Long streamwise correlations in $R_{uu}$ were found dominated by slower streamwise structures, since the streamwise streaks between zero crossings of 1500 wall units or longer occur more frequently for negative $u$ than positive $u$. In addition, correlation of $R_{ww}$ exhibits that the long streamwise slow-moving regions contain discrete zones of strong upwash over a large streamwise range, as might be present in packets of inclined hairpin vortices. At a wall-normal location over the log-law region ( $z/\delta = 0.5$ ), the correlation structures have a shorter scale in the streamwise direction and a broader scale in the spanwise direction. Correlation of $R_{uu}$ in the streamwise-spanwise plane is more elongated along the in-plane wall-normal direction than in the inclined cross-stream plane, in agreement with the presence of hairpin packets with a low-speed region lifting away from the wall.

## 2.3. Cross-Correlations

The cross-correlation coefficient is a measurement that tracks the fluctuations of two variables or sets of data relative to each other. The two sets of variables can be velocity, vorticity, shear stress, and other variables in studying turbulence. If no other explanation, cross-correlations in turbulence study are usually used to measure to two time series at one spatial point of turbulence fields.

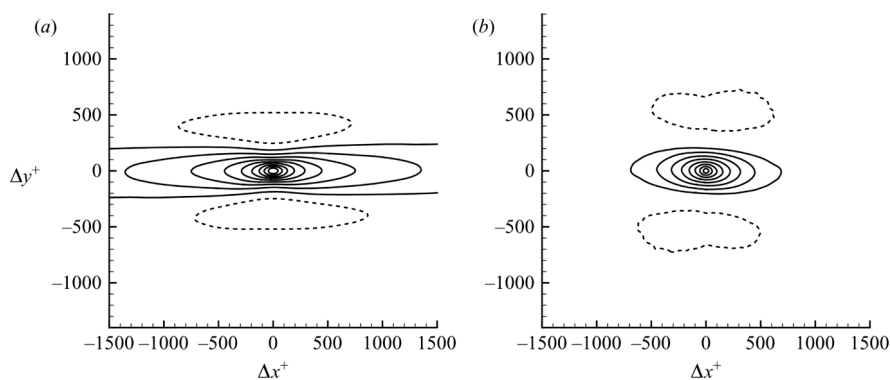

**Figure 1.** (a) $R_{uu}$ at the wall distance $z^+ = 92$, (b) $R_{uu}$ at the wall distance $z/\delta = 0.5$. The contour levels for $R_{uu}$ range from −0.1 to 1.0 with increment of 0.1 [32].

Klewicki (1989) derived the momentum equations in vorticity form, relations between velocity-vorticity correlations and gradients of the Reynolds stresses are established for a two-dimensional turbulent channel flow [33]. With these relations, the approximate formulas and recently obtained experimental data allow for unmeasured velocity-vorticity correlation profiles to be deduced [34]. The results indicate that the contributions to the $y$ gradient of $\overline{uv}$ were most equally shared between $\overline{v\omega_z}$ and $\overline{w\omega_y}$, while the contributions to the normal stresses are dominated by the $\overline{u\omega_z}$ term. In addition, the fact that the measured $\overline{vu\omega_z}/u_\tau^3$ profiles present little $Re_\theta$ variation implies that any $Re_\theta$ variation in the $\overline{v^2}$ and $\overline{w^2}$ gradients are caused by the variation of $Re_\theta$ in the correlation involving $\omega_x$.

Later, Klewicki, Falco, and Foss (1990) used time-resolved measurements of the spanwise vorticity component, $\omega_z$, to investigate the motions in the outer region of TBL [35]. The measurements were taken in very thick zero pressure gradient boundary layers ($Re_\theta = 1010 \sim 4850$) using a four-wire probe. An analysis of vorticity-based intermittency is reported near $y/\delta = 0.6$ and 0.85. The average intermittency is presented as a function of detector threshold level and position in the boundary layer. The spanwise vorticity signals were found to yield average intermittency values as large as previous intermittency studies. An analysis of $\omega_z$ event durations conditioned on the signal amplitude was also performed. The results suggested that for decreasing $Re_\theta$, regions of single-signed $\omega_z$ increase in size relative to the boundary layer thickness, but decrease in size when normalized by inner variables.

Two-point cross-correlations have been used to examine the influence of the complex boundaries on the near-wall vortices. Wang *et al.* (2019) applied the phase-resolved two-point cross-correlation between the wall shear stresses and streamwise vorticity to investigate numerically simulated turbulent flows over a wavy boundary with traveling-wave motion [36]. The correlation coefficient presents that near-wall streamwise vortices are closely associated with the wall shear stresses, and the magnitude of maximum correlation coefficient depends on the wave life time. As seen in **Figure 2**, the size of the correlation structures exhibits a remarkable variation with the wave phase for the case of slow traveling wave. When the wave is fast enough, the distribution of the correlation function recovers the manner similar with that in the flat-wall case. This method is expected to be applied to detecting near-wall streamwise vortices based on wall information in turbulent shear flows with complex boundaries.

Guo and Li (2010) performed a two-dimensional DNS of an incompressible two-dimensional turbulent channel with spectral method to examine the relation between wall shear stress and near-wall vortices [37]. Two-point correlation between the spanwise vorticity and the wall shear stress was calculated to explore the relation between these two variables. The results supported the notion that the wall shear stress is dominated by near-wall streamwise vortices.

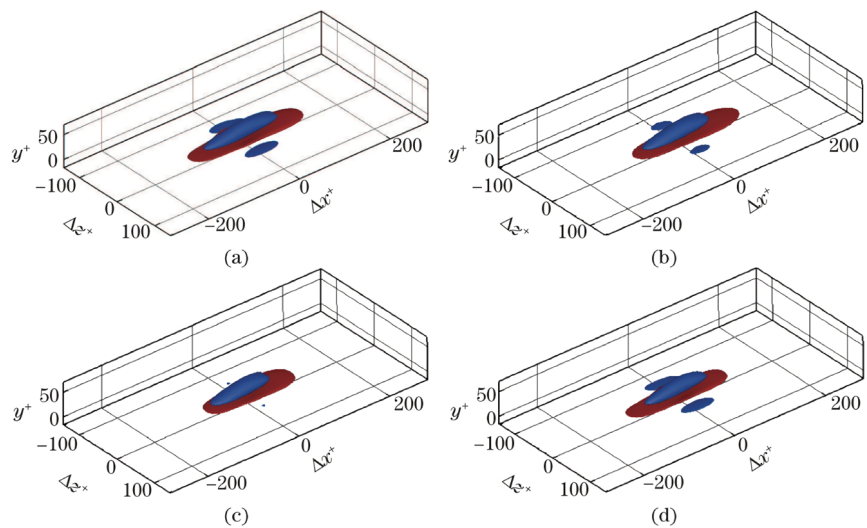

**Figure 2.** Iso-surfaces of the two-point cross-correlation coefficient between the streamwise vorticity and the spanwise wall shear stress for $R_{\tau_z \omega_x} = 0.15$ (red) and $-0.15$ (blue), (a) flat wall, (b) $c/U_m = 0$, (c) $c/U_m = 0.14$, and (d) $c/U_m = 1.4$ [36].

Some turbulent theories have also benefited from the cross-correlation technique. Deng *et al.* (2018) applied POD to two-dimensional PIV, together with a spatio-temporal coherence analysis to illustrate the downstream convection of the large-scale Q2/Q4 events, as the low-order signatures of the hairpin packets [38]. They applied the POD decomposition of the PIV-measured planar velocity fields, spatio-temporal correlation, and the combination of a POD-based low-order representation and two-point conditional correlation analysis to construct a statistical picture of vortex organization within a hairpin-packet.

In acoustic research, cross-correlations have been applied to seek the location of the sound source. Oguma, Yamagata, and Fujisawa (2013) examined an experimental method for detecting aerodynamic sound sources from a bluff body in turbulent flow [39]. Measurements were carried out in the sound field emitted from the circular cylinder at Reynolds number $Re = 4 \times 10^4$. With the help of the measured instantaneous velocity field using particle image velocimetry combined with the pressure Poisson equation, the pressure fluctuations around the circular cylinder were obtained. The sound pressure fluctuations were measured simultaneously by a microphone in the far field. The sound sources were then determined by cross-correlation of the pressure fluctuations on and around the flow field and the sound pressure fluctuations, consistent with the results in previous studies.

In geophysical research, understanding and quantifying the multiscale interactions between surface shear stress and velocity in the boundary layer is important in modeling TBL with proper boundary conditions. Venugopal (2003) conducted high-frequency measurements in a wind tunnel to identify dominant scales of interaction between wind velocity and shear stress through wavelet cross-correlation analysis [40]. The study provided way to estimate the linear correlation between shear stress and wind velocity at multiple scales and examine the reliability of boundary condition formulations in numerical models that compute shear stress as a linear function of wind velocity at the first vertical grid point.

# 3. Velocity-Vorticity Correlation Structures (VVCS)

## 3.1. Concept of VVCS

Correlation structures, such as hairpins, streamwise vortices, are believed to be the carriers of transporting turbulent energy. Several well-known methods have been prevailed for identifying the correlation structures in TBL, including the conditional sampling [41], the dynamic mode decomposition [42], the proper orthogonal decomposition (POD) [43], and the quadrant splitting method [44], to mention a few. However, the results from these methods can be easily influenced by the subjectiveness while determining the threshold value for the correlation analysis. Another fact about the methods in analyzing wall bounded turbulence is that enormous instantaneous flow fields are necessary for gaining a clear statistical description of turbulent structure. To handle the above issues, Chen *et al.* (2011, 2014) recently proposed a *velocity-vorticity correlation structure* (VVCS) method, with which the statistical CS can be easily extracted from turbulent shear flows [16] [45]. This method is meant to simplify the analysis of the wall-bounded and shear turbulence. In their study of compressible turbulent channel flows, the geometrical characteristics of the near-wall turbulence, such as the shape, the spanwise spacing, the streamwise length and the streak structures, were well achieved in a fairly straightforward way [16].

The concept of VVCS is constructed with high correlation regions in a field of two-point cross-correlation coefficient $R_{ij}$ ($i, j = 1, 2, 3$) of velocity component $u_i$ and vorticity component $\omega_j$, as defined in Equation (3). While velocity fluctuation $u_i$ at a reference location $y_r$, the regions of $\left| R_{ij} \left( y_r; x, y, z \right) \right| \geq R_0$ ($0 \leq R_0 \leq 1$) define a set of $VVCS_{ij}$. The analysis of compressible turbulent channel flows showed that the $VVCS_{ij}$ can accurately capture the geometrical features of near-wall CS, including spanwise spacing, the streamwise spacing and inclination angle of the quasi-streamwise vortices as well as the low-speed streaks [16].

## 3.2. Topology of VVCS

The VVCS illustrates the geometrical properties of three-dimensional vortical structure of wall-bounded turbulence. **Figure 3** shows the shape of velocity-vorticity correlation structure of $R_{11}$ in an incompressible channel flow at $Re_\tau = 180$. An inclined quadruple structure, with a pair of structures near the reference location $y_r^+$, and the other pair of structures attached to the wall. The correlations of the different velocity/vorticity components have quite different geometrical properties, as listed in **Table 1**. The overview of the correlation coefficient indicates that near-wall streamwise vortices are closely related to the near-wall shear stresses. In addition, the topology of a structure may vary with increasing the reference wall distance. For example, topological variation of $VVCS_{11}$ from quadrupole to dipole at $y_r^+ = 110$ was observed in compressible channel flow [16], as shown in **Figure 3**.

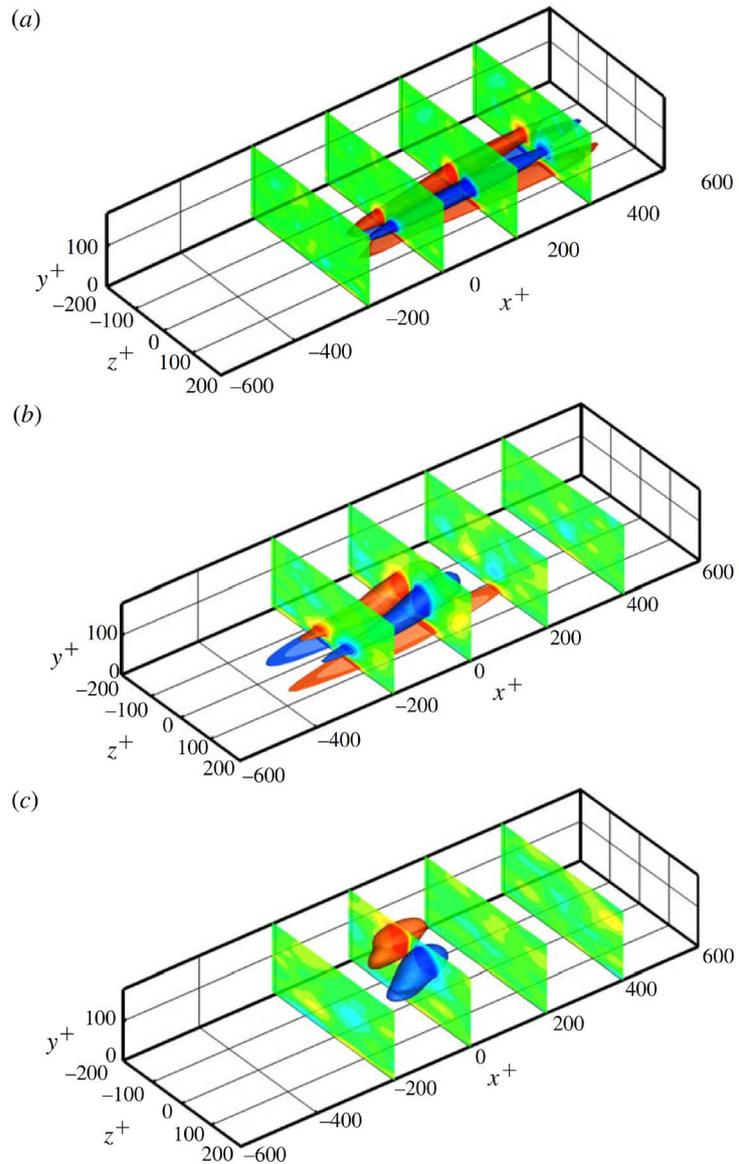

**Figure 3.** The iso-surface of the two-point cross-correlation coefficient for $R_{11}$ of an incompressible channel flow for $Re_\tau = 180$. The red surface is defined by the positive $R_{11} = -0.07$. The slices in the $y$-$z$ plane show distribution of $R_{11}$ with the spacing of threshold of $R_{11} = 0.07$, and the blue surface is defined by the negative threshold of otherwise. Note a topological change from (a) and (b) (four cigar-like elongated structures) $\Delta x^+ = 200$. The same threshold is used for identifying other VVCS unless mentioned to (c) (two blob-like structures): (a) $y_r^+ = 3.5$; (b) $y_r^+ = 59$; (c) $y_r^+ = 145$ [16].

**Table 1.** Topology of VVCS near the wall ( $y_r^+ < 45$ ). The subscripts $i, j (= 1, 2, 3)$ in $VVCS_{ij}$ denote $x$, $y$ and $z$ directions, respectively.

|   | $\omega_x$ | $\omega_y$ | $\omega_z$ |
|---|---|---|---|
| $x$ | Quadruple | Horizontal Dipole | Trident |
| $y$ | Quadruple | Horizontal Dipole | Trident |
| $z$ | Trident | Twins | Quadruple |

### 3.3. Spatial Relation between Structure Location and Reference Point

What is consistently found in VVCS studies is that the major contributor to the velocity fluctuations at reference location near the wall is the CS well above them. For instance, the near-wall convection velocity (propagation speed) is shown to be determined by distant vortices ($y_s^+ > 10$) [25]. Pei *et al.* (2013) proposed the concept of limiting structure of VVCS, defined as the structure at $y_r = 0$ [46]. The limiting VVCS as the streamwise vortical structure close to the wall was identified in numerical simulation data of compressible channel flows at the Mach number up to 3. A concrete Morkovin scaling summarizes all compressibility effects on the channel flow. Particularly, when the height and mean velocity of limiting VVCS are taken as the reference scales to normalize the variables, all geometrical measures in the spanwise and normal directions, along with the mean velocity and fluctuation profiles come to be Mach-number-independent. The results were validated by DNS data. The authors also reported similarity solution of the structure location $y_s$ as a function of the reference location $y_r$ based on the wall density. The geometrical similarity of the structure location consolidates the rationality of the semi-local transformation [47]. More importantly, the location and the thermodynamic properties related to the limiting VVCS can be used to describe the Mach number effects on compressible wall-bounded turbulence.

The Mach number effects on CS have attracted much attention in the studies on compressible turbulence. The Mach number dependence of length scale of CS in compressible channel flow at Mach numbers from 0.8 to 3.0 was investigated with VVCS [48]. Based on the notion of structure coordinate $(x_s, y_s, z_s)$, a set of scales characterizing the variation of topology structure at different Mach numbers were proposed in the study. It was shown that $VVCS_{11}$ and $VVCS_{12}$ are extensively stretched with increasing Mach number. Empirical relations of the length and spanwise spacing for the above two types of structures were suggested. It is noted that $VVCS_{12}$ can be considered as the statistical structure of low-speed streaks. The geometrical features of VVCS are consistent with the results of Coleman *et al.* (1995) [49]. This study also suggested that the relationship between the characteristic scales of VVCS and the Mach number should be considered in performing numerical simulation of compressible flows, to affirm whether the large-scales can be properly simulated in the computation domain.

The interpretation of the relationship between the statistical structure and the instantaneous structure is still an open question. Inclined "vortices" have been described as the eddies in boundary-layer eddies [50] and shear flows by sto-chastic estimation [13]. They are described as parts of local cascade structure and space-time correlations [24] [51]. However, it should be emphasized that the dimensions of the two-point velocity-vorticity correlations are much larger than those of individual vortices. The spacing between accompanying streamwise correlation structure (ASCS) of VVCS₁₁ linearly increases from 35 to 75 wall units, with increasing the reference wall distance from 0 to 150. It should be noted that the low-speed streaks and the cross-flow vortices must be associated, and the difference in their size makes it difficult to describe them as parts of a single vortex.

## 3.4. Physical Pictures of VVCS

The length of a streamwise structure, ASCS₁₁, decreases with increasing its wall-normal distance. It decreases from 500 to 150 wall units while the structure location $y_s^+$ moves from 15 to 150 wall units. There might be two reasons for these results: 1) the streamwise vortical motions becomes weak in the inertia region, and 2) three-dimensionality of the vortices are strengthened in the outer layer. Thus, a VVCS represents a statistical description of coherent motions. The VVCS analysis is a robust and efficient method for quantifying coherent motions in turbulent shear flows, and particularly suitable for extracting statistical geometrical measures using two-point simultaneous data from hotwire, particle image velocimetry, laser Doppler anemometry measurements and numerically simulated turbulent flows.

The rather simplified scheme of **Figure 4** is used to interpret the dynamics in the context of streamwise vortices with VVCS₁₁. If the statistical structures are considered as instantaneous structures, low-speed fluid ejects in between two pairs of counter-rotating streamwise vortices, *i.e.* ASCS, and the near-wall correlation structure (NWCS). This picture is consistent with the cause of generation of low-speed streaks. As mentioned in the previous section, the statistical structure may not exist in the instantaneous field, so the scenario of low-speed streak generation directly interpreted by statistical structures is not rigorous. However, comparing the results of the averaged structure and the energetic structures in the instantaneous turbulent flows might help understanding the formation of large-scale structures in turbulent flow. Accumulated evidence has suggested the existence of small-scale structures in the log-law layer and above, and they still contribute a significant part of turbulent kinematic energy.

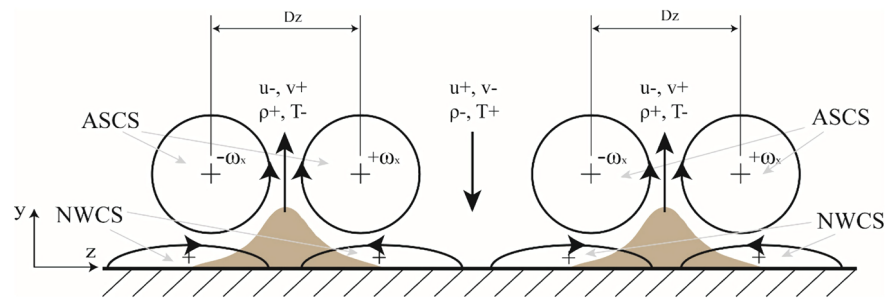

**Figure 4.** Schematic graph of the sub-structures of VVCS₁₁ and their dynamics near the cold wall.

### 3.5. Applications of VVCS

The VVCS is shown to reflect CS in different regions and thus can be used to illustrate vortical motions in the inner and outer regions, e.g. the lift-up process in TBL [52]. Farano *et al.* (2018) applied VVCS to extract CS from the DNS of turbulent channel flow and from the optimal perturbations at target time [53]. The VVCS was used to as the connection between the streamwise vortices and the velocity components. Particularly, the first component of the tensor, $R_{11}$, reflecting the correlation between streamwise vortices and streaks was utilized to identify the active lift-up regions. The quadruple structure characterized by elongated streaky structures was observed, indicating a strong correlation between flow at the reference point $y_r$ and the near-wall coherent motions. This property of VVCS$_{11}$ was found relevant to the lift-up mechanism, which was later confirmed by the presence of near wall streaks and streamwise counter-rotating vortices. With increasing $y_r$, the correlation becomes much weaker to the wall, in terms of a dipole shape. Variation of statistical structure was explained as the appearance of bulge structures which is closely linked to the head of hairpin vortices, frequently observed in the outer region. As found in the study, the optimal perturbation at $Re_\tau = 180$ is characterized by a very similar correlation to the TBLs. It should be seen that the onset of a bulge occurs at $y_r^+ = 100$, where hairpin vortices heads begin to lift up from the wall. In addition, for the case of the optimal perturbation at $Re_\tau = 180$, the NWCS of VVCS$_{11}$ exists even the reference wall distance over 100 wall units. The connection between the near-wall region and the outer one during the bursting process revealed by VVCS is attributed to the hairpin vortices in constituting the optimal perturbation.

The velocity-vorticity correlations were used to capture the CS in open channel turbulent flow, where visualization of structures is always challenging due to their multiscale and multi-layer natures. Bai *et al.* (2019) examined five types of methods for extracting CS in turbulent flow in the open channel, namely the Q-criterion, the vorticity, the Omega method, the VVCS method, and the Rortex method [54]. A DNS was performed to study the multi-layered flow structures under the free surface. The visualization results from the Q-criterion, the vorticity, the Omega method and the Rortex were calculated. The turbulent flow layers near the free surface were analyzed with corresponding anisotropy indices. They also extracted the VVCS in various turbulence layers. Among the five methods, only VVCS can straightforwardly achieve the geometry information of the CS throughout the whole domain of the open channel.

The VVCS analysis was used to reveal CS in a numerically simulated turbulent flow over a drag-reducing and a drag-increasing riblet configuration [55]. Impeding the spanwise movement of longitudinal vortices during the sweep events has been hypothesized to be a key mechanism for the turbulent drag reduction with the riblets. However, no direct evidence demonstrates that the riblets are capable to impede the spanwise motions in TBL. Three-dimensional two-point statistics was employed to quantify the interaction of the riblet surfaces with the coherent, energy-bearing eddy structures in the near-wall region.

Li and Liu (2019) measured geometrical properties of ASCS, e.g. the spanwise spacing $D_z$, as the reference wall distance $y_r$ varying from the wall to the logarithmic region in the DNS of turbulent boundary layer [55]. **Figure 5** shows that with the riblets the cores of the ASCS are closer to each other than those for flow over the smooth wall, and wider riblet spacing leads to more reduction of their spanwise spacings. This feature suggested that the riblets do impede the spanwise motions. Considering that the drag-increasing case is more effective to impede the spanwise motions, it may suggest that impeding the spanwise motions is not directly related with the drag reduction mechanism. Moreover, the linear function of $D_z^+\left(y_r^+\right)$ has a slope of 0.39 in comparison with that of 0.31 in compressible turbulent channel flow [16].

Reference [55] confirmed that the ASCS exists above the NWCS and inclines away from the wall, similar to the prediction of Townsend's attached eddy hypothesis [2]. The authors found that the streamwise vortices (ASCS$_{11}$) in the drag-reducing case are lifted up on the riblet tip, while in the drag-increasing case the streamwise vortices are implanted into the riblet cove. Since the region of high skin friction on the wall is mainly determined by the sweep motions of the streamwise vortices, the large area near the riblet tip occupied by ASCS$_{11}$ inducing high skin friction as being found in the drag-increasing case. Thus, the net drag is enhanced due to the increased drag near the riblet tip regardless of the drag reduction near the valley. In addition, the riblet surface tends to make the cores of streamwise vortices closer than that over the smooth wall, as shown in **Figure 5**. Wider riblet spacing is found to produce more drag reduction. In the cases with riblets the streamwise vortices have longer streamwise lengths, but their inclination angles do not change much. According to the transient growth (STG) theory, the NWCS of $VVCS_{11}$ is meant to be dominated by the internal shear layers (or vorticity sheets) and represents an averaged motion of near-wall vortices. In the cases with riblets, the spanwise size of NWCS is influenced by the inhomogeneous distribution of the wall surfaces. As $y_r$ increases, the NWCS remains attached to the wall until it disappears when $y_r^+ > 130$.

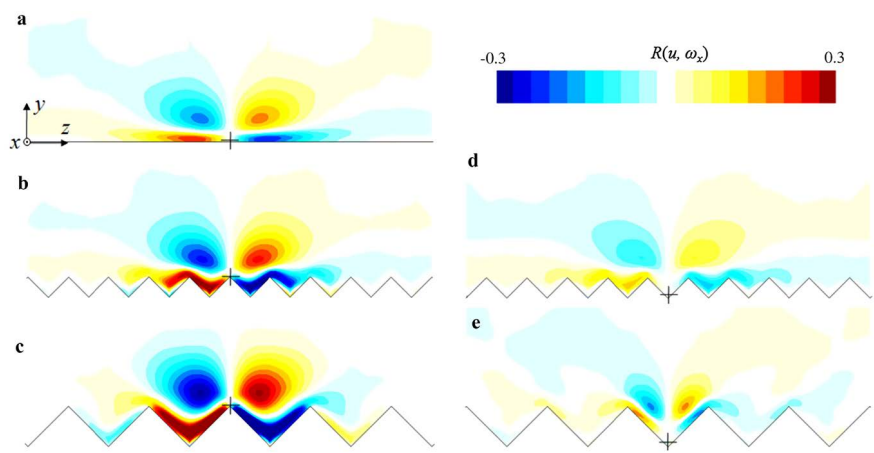

**Figure 5.** Contours of $R_{11}$ at the cross-flow plane where $x = x_r$ as $y_r$ approaches 0. The crosses "+" mark the positions of the reference points. (a) Baseline; (b) S-20, the reference point locates above the riblet tip; (c) S-40, above tip; (d) S-20, above valley; (e) S-40, above valley. Discussion of the figure is referred to [55].

The statistical CS in terms of VVCS quantitatively reveal the interaction between riblets and the near-wall vortical structures, enabling one to have an insight into the drag reduction mechanism. The lift-up of the ASCS in the riblet case was illustrated through examining the wall-normal distance of ASCS cores against the reference wall distance, $D_y(y_r)$, showing that $D_y$ in the riblet case at $y_r/\delta_{ref} < 0.1$ is higher than that in the smooth wall case, and no distinct variation of $D_y$ at $y_r/\delta_{ref} > 0.1$ was observed for all configurations of interest.

## 4. Concluding Remarks

The information in turbulent fields at two points and two times has extensively improved our understanding of the physics of turbulent flow in recent years. The major new achievement is that turbulence contains a remarkable degree of order. It can be expressed statistically in space-time correlations, in both Eulerian and Lagrangian frames of reference, which play a critical role in theories of turbulence. Approximately sixty years have passed since the earliest observations of coherent motions in TBL, but progress in applying these techniques to engineering is rather slow and the connection to a computational fluid model is still elusive. Recently, statistical structures based on velocity-velocity, velocity-vorticity, and other cross-correlations of other variables have shown their merits of quantifying coherent motions in turbulence. Such correlations as a type of experimental and numerical data analysis that have been, and will be, widely used in investigation of various types of turbulent flows.

## Acknowledgements

J. C. would like to show his warm thank to Professor Zhensu She who provided guidance and expertise that greatly assisted the VVCS study. J. C. also thanks Professor Fazle Hussain's insight that greatly supported the VVCS study. J. C. is grateful to Professor Yuhui Cao, Dr. Jie Pei, Dr. Xiaotian Shi, Dr. Ning Hu and Dr. Tiejin Wang for providing the reference data. This work was supported by the NNSF (grant numbers 11452002, 11172006, 10572004 and 11521091) and by MOST 973 project 2009CB724100.

## Conflicts of Interest

The author declares no conflicts of interest regarding the publication of this paper.

# Nomenclature

| | |
|---|---|
| $Ma$ | Mach number |
| $R_E$ | Eulerian correlation coefficient |
| $R_L$ | Lagrangian correlation coefficient |
| $R_{ij}$ | two-point correlation coefficient of $u_i$ and $\omega_j$ ( $i, j = 1, 2, 3$ ) |
| $Re$ | Reynolds number |
| $Re_\tau$ | Reynolds number defined with the friction velocity |
| $Re_\theta$ | Reynolds number defined with the momentum loss thickness of turbulent boundary layer |
| $t$ | time (s) |
| $U$ | mean velocity (m·s⁻¹) |
| $U_c$ | propagation speed (m·s⁻¹) |
| $(u_1, u_2, u_3)$ or $(u, v, w)$ | velocity components in the streamwise, wall normal and spanwise directions, respectively (m·s⁻¹) |
| $(x_1, x_2, x_3)$ or $(x, y, z)$ | coordinates in the streamwise, wall normal and spanwise directions, respectively (m) |
| $\delta$ | thickness of turbulent boundary layer (m) |
| $\nu$ | kinematic viscosity (m²·s⁻¹) |
| $\phi$ or $\psi$ | fluctuations of a variable or component |
| $(\omega_1, \omega_2, \omega_3)$ or $(\omega_x, \omega_y, \omega_z)$ | vorticity components in the streamwise, wall normal and spanwise directions, respectively (s⁻¹) |

### Subscripts

| | |
|---|---|
| $i$ or $j$ | coordinate indices, ( $i, j = 1, 2, 3$ ) |
| $r$ | reference point |
| $x, y, z$ | coordinate indices for the streamwise, wall normal and spanwise directions, also represented by 1, 2 and 3, respectively |

### Superscripts

| | |
|---|---|
| + | normalized with wall units |